%% file: spinbond_arxiv.tex
\def\AUTHORa{Silvio Capobianco
 \thanks{This research was supported by the European Regional Development Fund
(ERDF) through the Estonian Center of Excellence in Computer Science (EXCS), and by the Estonian Science Foundation under grant \#7520.}
 }
 \def\AUTHORb{Tommaso Toffoli}
 \def\TITLE{Can anything from Noether's theorem
\\be salvaged for discrete dynamical systems?}
 \long\def\ABSTRACT{The dynamics of a physical system is linked to its
phase-space geometry by Noether's theorem, which holds under standard
hypotheses including continuity. Does an analogous theorem hold for discrete
systems?  As a testbed, we take the Ising spin model with both ferromagnetic
and antiferromagnetic bonds. We show that---and why---energy not only acts as a
generator of the dynamics for this family of systems, but is also conserved
when the dynamics is time-invariant.}

\documentclass{article}
 \usepackage{amsmath,amsfonts,url,graphicx}
 \input pack10.sty
 \input ttpaper

 \input latex+4

 \def\squeeze{\itemsep0pt\parsep0pt\parskip2pt}

\begin{document}
 \author{\AUTHORa\ and \AUTHORb}
 \title{\TITLE}
 \abstract{\ABSTRACT}

\title{\TITLE}

\maketitle

\Sect[intro]{Introduction}

In the last three hundred years, analytical mechanics has turned up a wealth of
profound concepts and powerful methods for dealing with the dynamical systems
of \emph{physics}. Can these results---of which Noether's Theorem is one of the
best known---be applied to the dynamics of \emph{information systems}, which
today make up such a large part of our culture, and to what extent?

Here one may anticipate two difficulties. On one hand, the above results
take advantage of specific aspects of physics and may not be valid outside that
territory. On the other, the methods of analytical dynamics are intimately tied
to the calculus of the \emph{continuum}, and may fail when applied to
\emph{discrete} systems such as automata and networks.

In sum, is analytical dynamics just a ``bag of tricks'' for doing continuum
physics, or does it provide templates for tools useful in a more general
context? Does the insight that it offers transcend the distinction between
physical and man-made systems (cf.~\cite{tyagi}) and that between discrete and
continuous ones?

\medskip

The prototypical dynamical system that we all know and love, namely,
\emph{physics}, displays a striking property. Its explicit evolution law has
the form of a lookup table that to each possible \emph{state} of the
system---say, a $2n$-tuple of real variables (or a \emph{vector}) consisting of
$n$ positions and $n$ momenta (where $n$ is the number of degrees of
freedom)---associates another $2n$-tuple---its ``next state.''  For physical
systems, it turns out that this table can be ``compressed'' into one that to
each state associates a \emph{single real variable}---the \emph{energy} of that
state. The first table explicitly expresses the dynamics as a \emph{vectorial}
function, while the second implicitly expresses the dynamics through a
\emph{scalar} function. This ``compression'' is \emph{nondestructive}, that is,
the full table can be reconstructed from the compressed table.  Both the
$2n$-to-1 encoding and the converse 1-to-$2n$ encoding are achieved by simple,
standardized algorithms (in analogy with ``zipping'' and ``unzipping'' a file).
For this reason, the \emph{energy function}, that is, the above energy lookup
table, is said to be the \emph{generator} of the dynamics, in the sense that
plugging this function into a generic kinematic scheme (such as Hamilton's
canonical equations) allows one to explicitly reconstruct a specific dynamics.

The usual way to derive the above property of energy relies on the fact that a
physical system (both in classical and quantum mechanics) is imagined to evolve
\emph{continuously} in time. However, in order to have this property, being
time-continuous is not enough---for this, the dynamics must have much
additional structure, namely, \emph{symplectic}, in classical mechanics, and
\emph{unitary}, in quantum.

As we generalize the concept of dynamical system, in particular to systems
that, like automata of all kinds, have \emph{discrete} state variables and
evolve \emph{discretely} in time, the question arises whether an energy-like
quantity can be retained at all, and, if so, under what conditions. The first
question we'll ask here is whether a cellular automaton can have something that
can rightfully be called ``energy,'' what conditions it would have to satisfy
for that, and what it would look like. We'll then ask whether and
to what extent the method proposed by Noether for deriving \emph{conserved
quantities} from the \emph{symmetries} of the dynamics of a continuous systems
may be adapted to discrete systems.

\Sect[conserve]{Energy conservation}

We've all heard of the principle of {\sl conservation of energy}: In an
isolated physical system, no matter what transformations may take place within
it, there is a quantity, called {\sl energy}, that remains conserved.  Our
belief in this principle is so strong that, whenever a violation of energy
conservation is observed, our first reflex is not to question the principle,
but to look more carefully into the experimental evidence and ask whether
something may be escaping us.
 To defend the reputation of energy as a bona fide conserved quantity against
new incriminating evidence, now and then physicists are even willing to stretch
energy's very definition (see Feynman's peerless
parable\cite{feynman63energy}).

How far shall we be able to retain a viable concept of energy as our conception
of ``dynamical system'' keeps widening? Can energy be conserved in a
\emph{cellular automaton}? (cf.~\cite{boykett08})
Conserved or not, should an energy be expected to be found
\emph{at all} in the latter? Is the idea of ``energy'' meaningful for a
\emph{Turing machine} or a \emph{finite automaton}? In sum, what features does
a dynamical system need to exhibit in order to possess a recognizable
generalization of energy?
\emph{Why}
should energy be conserved? 

\Sect[noether]{Noether's theorem}

Nearly a century ago, Emmy Noether formulated a principle of
great generality and beauty, commonly known as {\sl Noether's
Theorem}\cite{noether18,wikipedia11noether}, which we shall state for a moment
just in headline form, ignoring the fine print: \emph{``To every
symmetry of a dynamical system there corresponds a conserved quantity.''}
According to this principle, the conservation of energy (a \emph{state
variable}) would be the consequence of the constancy of a system's dynamical
laws, that is, of the invariance in time of the system's \emph{structural
parameters}.

Noether's theorem plays a unifying role in our understanding of mechanics;
according to \cite{wikipedia11noether}, it ``has become a fundamental tool of
modern theoretical physics.'' Since it can take advantage of many kinds of
symmetries, it is very productive: for instance, it predicts that, if a
system's laws are rotationally invariant, then, just because of that
structural invariance, there will be a specific state variable that is
conserved, namely, \emph{angular momentum}.

\medskip

However, when we get down to reading Noether's theorem's fine print,
we discover a number of restrictive provisos. A fuller version of the contract
would spell ``To every \emph{one-parameter}, \emph{continuous group} of
symmetries of a \emph{Lagrangian} dynamical system there corresponds a
\emph{scalar}, \emph{real-valued} conserved quantity.'' Thus, a \emph{discrete}
group of symmetries, such as the invariance of Conway's ``game of {\sc life}''
under quarter-turn rotations of the orthogonal grid, won't do (bye-bye angular
momentum?). Similarly, even though the dynamics of {\sc life} is the same at
every step, and thus \emph{time-invariant}, it is not \emph{invertible} (and
thus only forms a semigroup, not a group), and in any event the game advances
in discrete steps, instead of along a one-dimensional continuum (bye-bye
energy?).

\Sect[ising]{The Ising playground}

The best way to identify the issues at play is to examine a specific
class of models, and then map the conceptual question of our
investigation to concrete features of a model. For our purposes, we
shall choose a standard cellular-automaton realization of
the \emph{Ising spin model}
(cf.~\cite{bach07,bennett88bond,pomeau84}). This consists of a regular
array of elementary magnets, or \emph{spins}, that can assume one of
two orientations---`up' and `down'.

Spins interact with one another.  In the
Ising model stylization of physics, long-range forces are ignored and
short-range ones represented by the interaction of a spin with only its
immediate neighbors: its four first neighbors, in a
2D
orthogonal
array. In the simplest case, these four \emph{bonds} (or ``couplings'') are all
of a \emph{ferromagnetic} nature, that is, they behave like rubber bands that
are relaxed when the spins at the two ends of a band point in the same
direction (``parallel''), and stretched, when in the opposite direction
(``anti-parallel'').

Different kinds of dynamics may be used for the Ising model;
in order
to have a unifying criterion for prescribing a dynamics, one starts with
assigning one unit of a notional ``potential energy'' to stressed bonds, and
zero to unstressed ones. The class of dynamics we choose here, the so-called
\emph{microcanonical Ising model}, are strictly deterministic and invertible:
on a given step, a spin will \emph{flip} (that is, reverse its orientation)
\emph{if and only if} doing so will leave the sum of the potential energies of
the four surrounding bonds \emph{unchanged}.  Note that, in the ferromagnetic
case, this happens when two of the neighbors are parallel to the given spin and
the other two anti-parallel.
  In this case, in fact, we start with two stressed bonds. After the flipping
of a given spin, its parallel neighbors will have turned anti-parallel and vice
versa, so that the two stretched bonds relax at the same time as the two
relaxed ones stretch, and consequently the overall potential energy stored in
these bonds is two units both \emph{before} and \emph{after} the step. In all
other cases, since no flip occurs, potential energy is conserved trivially.

A final stipulation. The above rule would become inconsistent if one attempted
to update two adjacent spins \emph{at the same time}. In fact, if \emph{both}
spins are instructed to flip, the assumption on the part of either spin that
the shared bond's stress status will thereupon be complemented \emph{fails} (as
its stress status will remain unchanged), and energy is no longer conserved.
  A standard solution is to treat the array
of spins as a checkerboard consisting of two intermeshed subarrays called
\emph{even} and \emph{odd} (the red and black squares of the board) according
to the parity of $x+y$, where $x$ and $y$ are the spatial coordinate of a site.
A system \emph{state} $s$ consists of an \emph{ordered pair} $\langle
q_A,q_B\rangle$ of \emph{configurations} of different parity (\ie\ one even and
one odd, in either order). Together, an even and an odd configuration fill up
the whole array; however, their order in the pair is relevant, that is, state
 $\langle q_A,q_B\rangle$ is distinct from $\langle q_B,q_A\rangle$. Depending
on the context, it may be convenient to call the two configurations of a pair
either the ``past'' and the ``present'' or, alternatively, the ``present'' and
the ``future'' (the latter abbreviated, when subscript, as ``pres'' and
``futr'').

The Ising dynamics is a second-order recurrence relation of the form
 \Eq[2nd]{
  q_{x,y}^{t-1}\oplus f(q_{x-1,y}^t,q_{x+1,y}^t,q_{x,y-1}^t,q_{x,y+1}^t) =
q_{x,y}^{t+1}.
 }
 \noindent This is a \emph{conditional permutation}: a spin $q_{x,y}$ in the
\emph{past} (time $t-1$) is mapped into a spin $q_{x,y}$ in the \emph{future}
(time $t+1$) by {\sc xor}ing it with an \emph{enable} bit---the output of the
binary function $f$ when given as arguments the \emph{present} values (time
$t$) of the first-neighbor spins $q_{x-1,y},q_{x+1,y},q_{x,y-1},q_{x,y+1}$,
which belong to the complementary subarray and thus affect the updating of
$q_{x,y}$ but are not affected it. Depending on the value of the enable bit,
this updating either leaves a spin unchanged or complements it---both
invertible operations. The above ``shifting-frame'' mechanism---of the form
$\langle q_\text{past},q_\text{pres}\rangle\mapsto\langle
q_\text{pres},q_\text{futr}\rangle$, where the right item $q_\text{pres}$ of
the old state shifts to the left unchanged, pushing out the left item
$q_\text{past}$, and a new item $q_\text{futr}$, computed according to
\eq{2nd}, shifts in from the right---is called in numerical analysis
\emph{leapfrog updating}.

\medskip

The macroscopic behavior of this kind of model is trivial in one dimension;
conceptually productive in two dimensions, as it provides by very simple means
an insight into the physics of phase transitions;
 and increasingly rich and challenging in more dimensions.

\Sect[why]{Why call it energy?}

Setting physical motivation aside, we have in the Ising spin model an abstract
symbolic dynamics consisting of a two-dimensional array over the binary
alphabet $\{0,1\}$ (for what was `down' and `up'; it will still be convenient
to call ``spin'' the state variable at each site). We distinguish between even
and odd sites according to the parity of the sum $x+y$ of their coordinates.
We update even and odd sites on alternating steps according to the rule (stated
the rule in a way that obviously generalizes to any number of dimensions) that

{\narrower\noindent ``A spin $u$ at $\langle x,y\rangle$ flips
(\ie\ complements) \emph{if and only if} its first neighbors (those at $\langle
x\pm1,y\rangle$ and $\langle x,y\pm1\rangle$) are \emph{evenly divided} between
state $u$ and its complement $\overline u$.''\par}

\noindent Note that we had no use for an ``energy.'' However, if as an
afterthought we choose to call \emph{energy} a distributed quantity whose
value, for each \emph{pair} of adjacent first-neighbor spins, is 0 if these two
spins are equal and 1 if different, then it follows that energy is
\emph{conserved} by the above dynamics. Also note that this energy is but the
\emph{length of the boundary} between `up' and `down' domains. While the
\emph{magnetization}---number of spins up minus number of spins down---may
change with time, that length, and thus the energy, remains constant. Is this
use of the term ``energy'' just physics nostalgia---like calling a British
colony ``New Hampshire''? Or is there something more to it?

\medskip

For most people, energy is a fungible resource of the form $E=T+U$, that is,
stored either in the inertia of a moving object (``kinetic energy'' $T$) or in
the state of compression of an elastic medium such as a spring (``potential
energy'' $U$).
Such a concrete presentation of energy will do
for many practical applications.

But if you look at what energy is \emph{really} supposed to be in physics,
you'll discover an abstract quantity that is totally unconcerned with the
\emph{nature} of the objects and materials that make up a dynamical system, and
instead totally absorbed with the topological structure of the system's
\emph{trajectories}. In fact, the total \emph{energy} of a system may be
defined as
{\abovedisplayskip0pt\belowdisplayskip0pt
 \Eq[energy]{\parbox{3in}{
  \begin{enumerate}\squeeze
  \item A \emph{real-valued function} of the system's \emph{state},
  \item that is \emph{additive},
  \item and is a \emph{generator of the dynamics}.
  \end{enumerate}}
 }
}
\noindent Incidentally, there is no mention of ``energy conservation'' in all
this.

\begin{fineprint}\noindent{\sc Note.}\quad
 We wish we could give a precise reference for the above definition. In fact,
here we rely on disparate loose hints culled and integrated with some effort
from the literature.  Physicists are supposed to know what energy is, and they
don't need---perhaps have good reasons not to \emph{want}---a final
definition (cf.\ the Feynman parable mentioned in \sect{conserve}). So, not
only college textbooks, but even such beloved conceptually-minded references as
Arnold\cite{arnold10methods}, Goldstein\cite{goldstein01mechanics}, and
Lanczos\cite{lanczos86variat} proceed by giving \emph{examples} of energy in
increasingly elaborate contexts, and then bringing one's attention to
increasingly more abstract \emph{properties} of energy. Definitions---hardly
ever.\end{fineprint}

\medskip 

Let's verify that the ``energy'' of the Ising model agrees with our abstract
definition. In order to do so, we'll be forced to re-examine the definition and
pore over the fine print that by tacit agreement stands behind such terms as
``state,'' ``additive,'' and ``generator.''

\Subsect[additive]{Additivity}

As we've seen, what we proposed to call Ising ``energy'' of a state is simply the
\emph{number of streteched bonds} in that state. Incidentally, since the Ising
model consist of an \emph{infinite} array of sites, the state of the system
consists of an infinite array of spins, and virtually any state will have
\emph{infinite} energy. For a function to be \emph{additive}, here the
physicist means that (a) it is meaningful to partition the system into separate
subsystems such that each of these subsytems has its own well-defined state, and
the global state is the composition, or Cartesian product, of the states of
these subsystems; (b) the function is defined on the state of each of these
subsystems, and (c) its value for the whole system is the sum of its values for
the individual subsystems.

If a system could be partitioned into subsystems or ``blocks'' that are totally
independent of one another, each one having a well-defined energy independent
of the state of the other blocks, additivity would be trivial to achieve---just
\emph{define} the energy of the whole system as the sum of the energy of its
subsystems---but by the same token would be a \emph{vacuous} concept.  In
general, however, there will be some residual coupling between blocks, and even
if one knew the precise state of a block one would be able to assign an energy
to it only to within an upper bound and a lower bound, which reflect the lack
of knowledge of the block's environment's state. Fortunately, in many cases,
when interactions are \emph{local} and the blocks are appropriately chosen,
this uncertainty is roughly proportional to the length of a block's
\emph{boundary}---where it interfaces with its environment, and thus the
\emph{relative} uncertainty is bounded above by the boundary/volume ratio for a
block, decreases with block size, and vanishes in the limit of arbitrarily
large blocks.  This is what ``additivity'' means to the physicist, and it is
consistent with the approach taken by topological dynamics.

In the Ising model, sites interact with one another through an unbroken mesh of
couplings; there are no strictly independent subsystems.  However, the
stipulation made at the end of \sect{ising}, whereby the odd mesh is kept
constant when the even mesh is updated (and vice versa)---and thus effectively
acts as a parameter rather than a state-variable---enforces a discipline by
which energy variations corresponding to local variations of a configuration
can be determined \emph{exactly}. We shall see this at the end of
\sect{generate} and use it to full advantage in \sect{spacetime}.

\Subsect[generate]{Generator of the dynamics}
 
A \emph{generator of the dynamics} is a function of the system's state through
the knowledge of which one can reconstruct the system's dynamics
 in an explicit, vectorial form (\cf\ \sect{intro})---of course only \emph{up
to an isomorphism}. It is understood that this reconstruction should be
achievable by an algorithm given once-and-for-all,
 independent of the specific system (otherwise, one could ``cheat'' and hide
information about the system in the algorithm itself---big deal!).  Suppose the
function is a ``Hamiltonian'' that for each state $\langle q,p\rangle = \langle
q_1,\dots,q_n,p_1,\dots,p_n\rangle$ yields its energy $H(q,p)$.  Our goal is to
determine where any state $\langle q,p\rangle$ will go. 

For a continuous dynamics as we imagine in the present example one would want
the instantaneous ``direction and rate'' at which the state progresses, that
is, the $2n$ derivatives $\langle \dot q=\frac{dq}{dt},\dot
p=\frac{dp}{dt}\rangle$, In other words, we want to arrive at \emph{vectorial}
mapping, of the form
 \Eq[rate]{
 \langle q,p\rangle\mapsto\langle \dot q,\dot p\rangle,
 }
 even though mere evaluation of the function at $\langle q,p\rangle$ will
perform the mapping $\langle q,p\rangle\mapsto H(q,p)$ and thus only give us a
a scalar.
 However, if we performed \emph{repeated} samplings of $H$ in the vicinity of
$\langle q,p\rangle$ (after all, we have the entire function $H$ and we can
apply it as many time as we want) so as to get a ``sense of direction,'' or,
equivalently, in the continuum case, if we could get the \emph{derivatives} of
$H$ at that point, we would get more information associated with that point
than its mere energy $H(q,p)$.
 In Hamiltonian mechanics, mapping \eq{rate} is provided by the standard
\emph{canonical equations}
 \Eq[canon]{
  \dot q = \frac{\partial H}{\partial p};\quad \dot p = -\frac{\partial
  H}{\partial q}.
 }
 \noindent Note that, since $q$ and $p$ are actually abbreviations for
$n$-tuples, here we have $2n$ equations yielding a \emph{vector} consisting of
$n$ pairs $\langle\dot q_i,\dot p_i\rangle,\ (i=1,\dots,n)$.

\medskip

It will presently become obvious that in the Ising model the energy function
\emph{does} generate the dynamics---and by means of a very simple recipe.  In a
\emph{discrete} Hamiltonian dynamics, a state is no longer a
``position/momentum'' pair $\langle q,p\rangle$ as in the continuous case, but
an ordered \emph{pair of configurations} $\langle q_0,q_1\rangle$, as explained
at the end of \sect{ising}.  The rules of the game are that one may propose
\emph{any} such state, \emph{as many times as one wishes}, and each time get as
an answer the energy $H(q_\text{past},q_\text{pres})$ for that state. Isn't
this number, in the range $\{0,1,2,\dots,\infty\}$ (and almost always $\infty$)
much too little to determine the infinite $q_\text{futr}$ configuration?  The
answer lies in the principle of \emph{virtual displacements}. That is, given
$q_\text{past}$ and $q_\text{pres}$, one may propose \emph{any conceivable
candidate} for $q_\text{futr}$, ask for the energy of the resulting new state
$\langle q_\text{pres},q_\text{futr}\rangle$, compare it with the old energy,
and on the basis of this comparison decide whether to accept or reject the
candidate. Just as in the ``hot, hot, cold, cold'' game, in the game of
``battleships''---or, for that matter, in biological evolution---one is never
explicitly shown the target.  Instead, one must describe an object, and will
only be told (``yes,'' ``no'') whether that is target, or possibly given an
intermediate scalar value (``hot'', ``warm'', etc.) that indicates \emph{how
far} one's proposal is from the target. In principle, one can arrive at the
solution by going through all possible candidates. The point of the game is to
speed up the search by dynamically arranging the proposals so that the feedback
will give one a ``sense of direction.''

In the Ising case, we want to find a $q_\text{futr}$ such that the energy
difference $\Delta
E=H(q_\text{pres},q_\text{futr})-H(q_\text{past},q_\text{pres})$ is
zero not only globally but also \emph{locally}, that is, at any scale down to
the radius of the first-neighbor neighborhood (the range of interspin
``interactions''). Forget about getting finite energies by taking finite
configurations by blocking (\sect{additive}); just take the 32 \emph{infinite}
configurations of the form

\kern-10pt
 {\small
 \Eq[stencil]{
 \begin{matrix}
   & \vdots & \vdots & \vdots & \vdots & \vdots & \\
  \cdots & 0 & 0 & 0 & 0 & 0 & \cdots\\
  \cdots & 0 & 0 & n & 0 & 0 & \cdots\\
  \cdots & 0 & w & c & e & 0 & \cdots\\
  \cdots & 0 & 0 & s & 0 & 0 & \cdots\\
  \cdots & 0 & 0 & 0 & 0 & 0 & \cdots\\
   & \vdots & \vdots & \vdots & \vdots & \vdots &
 \end{matrix}
 }
 }%end small
\kern-10pt

\noindent with $c$ denoting a ``center cell'' in the past and $n,s,w,e$
denoting its four neighbors in the present. This infinite state will have a
finite energy ranging from 0 through 4. Let's make a table of energy
for the 32 possible collective values of those  five sites,

\kern-10pt
 {\footnotesize
 \Eq[flip]{
 \begin{tabular}{@{}c@{\,}c|c@{\,}c@{\ \ }c@{}}
 $c$ & $nswe$ & $H$ & $f$ & $c'$\\\hline
 0 & 0000 & 0 & 0 & 0\\
 0 & 0001 & 1 & 0 & 0\\
 0 & 0010 & 1 & 0 & 0\\
 0 & 0011 & 2 & 1 & 1\\
 0 & 0100 & 1 & 0 & 0\\
 0 & 0101 & 2 & 1 & 1\\
 0 & 0110 & 2 & 1 & 1\\
 0 & 0111 & 3 & 0 & 0\\
 0 & 1000 & 1 & 0 & 0\\
 0 & 1001 & 2 & 1 & 1\\
 0 & 1010 & 2 & 1 & 1\\
 0 & 1011 & 3 & 0 & 0\\
 0 & 1100 & 2 & 1 & 1\\
 0 & 1101 & 3 & 0 & 0\\
 0 & 1110 & 3 & 0 & 0\\
 0 & 1111 & 4 & 0 & 0
 \end{tabular}\quad\quad
 \begin{tabular}{@{}c@{\,}c|c@{\,}c@{\ \ }c@{}}
 $c$ & $nswe$ & $H$ & $f$ & $c'$\\\hline
 1 & 0000 & 4 & 0 & 1\\
 1 & 0001 & 3 & 0 & 1\\
 1 & 0010 & 3 & 0 & 1\\
 1 & 0011 & 2 & 1 & 0\\
 1 & 0100 & 3 & 0 & 1\\
 1 & 0101 & 2 & 1 & 0\\
 1 & 0110 & 2 & 1 & 0\\
 1 & 0111 & 1 & 0 & 1\\
 1 & 1000 & 3 & 0 & 1\\
 1 & 1001 & 2 & 1 & 0\\
 1 & 1010 & 2 & 1 & 0\\
 1 & 1011 & 1 & 0 & 1\\
 1 & 1100 & 2 & 1 & 0\\
 1 & 1101 & 1 & 0 & 1\\
 1 & 1110 & 1 & 0 & 1\\
 1 & 1111 & 0 & 0 & 1
 \end{tabular}
 }
 }% end footnotesize
\kern-20pt

\noindent where $H$ is the ``Hamiltonian energy'' of the Ising model for the
entire global state \eq{stencil}, $f$ the ``conditioning'' function that
appears in \eq{2nd}, and $c'$, as we shall see in a moment, the effective
future value of the center site. As we recall from \sect{ising}, the Ising
dynamics is ``flip if and only if that leaves the energy of the neighborhood
unchanged.'' Therefore, for any state of the form \eq{stencil}, we may propose
as candidates for the new state either (a) the same state, \ie\ with $c'=c$, or
(b) the one where $c$ has flipped, \ie\ where $c' = \overline c$. By using
table \eq{flip} \emph{two} times, \ie\ for the two $c'$ candidates $c$ and
$\overline c$, we see that the energy is (obviously) unchanged in the first
case, and \emph{may} have changed in the second case. If the energy is the
\emph{same} in both cases, then the \emph{actual} value for $c'$ (that is, the
one prescribed by the Ising dynamics) will be the default value $c$; otherwise,
it will be $\overline c$. This is the value given in
column $c'$ of \eq{flip}.

Finally, we must show that this new value $c'$, given the old value of the
entire neighborhood
  $\vbox{\hbox{\def\arraystretch{0.55}\footnotesize$\begin{array}{@{}c@{}c@{}c@{}}
& n\\w&c&e\\&s\end{array}$}}$, is independent of all the others sites of the
global state---those assigned 0 in global state\eq{stencil}. In fact, by the
definition, in the Ising model, of energy as a \emph{sum over bonds} (or
adjacent-spin pairs), the energy spanning, say, the ``extended patch''
 $\vbox{\hbox{\def\arraystretch{0.55}\footnotesize$\begin{array}{@{}c@{}c@{}c@{}c@{}}&&u\\&n&a&r\\w&c&e\\&s\end{array}$}}$
 is the sum of the energies spanning the two elementary patches
 $\vbox{\hbox{\def\arraystretch{0.55}\footnotesize$\begin{array}{@{}c@{}c@{}c@{}}
& n\\w&c&e\\&s\end{array}$}}$ and
 $\vbox{\hbox{\def\arraystretch{0.55}\footnotesize$\begin{array}{@{}c@{}c@{}c@{}}
& u\\n&a&r\\&e\end{array}$}}$; in other words, energy is \emph{strictly
additive}, and the \emph{local dynamics} table \eq{flip} can be applied
separately to every elementary patch.

\medskip

Thus, the ``energy'' of the Ising model satisfies the abstract,
``model-independent'' definition of \emph{energy} given in \eq{energy}. \dots\ Or
\emph{does} it?

\Sect[bar]{Raising the kinematics bar}

Indeed the Ising energy (a) is a real-valued function of the system's state;
(b) it is additive, since the alternating updating scheme guarantees that the
accounting of energy can be done independently on every elementary patch; and
(c) a simple rule allows us to reconstruct the dynamics by sampling the energy
function in the (topological) neighborhood of a state.

However, point (c) is in this case \emph{vacuous}, since here we are dealing
with a \emph{single} system---the Ising model---and there is no indication of
how one would go about generating the dynamics of \emph{other} systems using
other energy functions.  
  Consider a telegraph channel with a single input key, pressing which will
invariably transmit the text of the {\sl Kamasutra}. In spite of the length and
the intrinsic interest of this text, the capacity of the channel is obviously
$\log 1=0$ bits. In a similar way, the information contents of $H_\text{ising}$
\emph{as a function}, for the sake of specifying a dynamics, is zero, because
it is a single ``key'': the Ising dynamics is the only dynamics we have in
mind, it is known, and we might as well have stored a copy of it at the
channel's receiving end.  What one needs is a generic function scheme---a look-up
table ``blank form,'' as it were---which, when filled with specific contents
and processed by an algorithm given once and for all, will allow one to
describe any one of an agreed upon class of dynamical systems.

\medskip

This issue is addressed by physicists by distinguishing between ``dynamics''
and ``kinematics,'' where the latter term is used in a somewhat specialized
sense. That is, a \emph{kinematics} is a framework---a set of rules---that
characterizes a whole class of dynamical systems of a similar type or
``texture,'' while a \emph{dynamics} is any one of the systems sharing that
kinematics. Until we place the Ising model within a larger class of models, and
show the existence of an energy-function-valued \emph{variable}---whose values
will be specific energy functions each of which will in turn allow a fixed
algorithm to generate the different dynamics of that class---until then we will
not have a bona fide \emph{generator of the dynamics}.

 There is a natural and a very simple generalization of the Ising model (used,
specifically, in the modeling of \emph{spin glasses}) that yields a class of
\emph{sixteen} dynamics. We'll show that for that class the energy function
plays a nonvacuous role as a generator of the dynamics.

\medskip

In \sect{ising}, we called \emph{ferromagnetic} a bond between adjacent spins
that is unstressed when the two spins are parallel and stressed when antiparallel
(``type 0'' bond). We now introduce a bond of ``type 1,'' or
\emph{antiferromagnetic}, which is stressed when the spins are \emph{parallel},
and unstressed otherwise. With bonds of two kinds thus available, and four
bonds surrounding a spin, there are $2^4=16$ possible \emph{structural
environments} by which a spin can find itself surrounded.  We shall retain our
definition of \emph{Ising energy} as the sum of the energy of the bonds,
recalling only that the stress $u\stackrel b\longleftrightarrow v$ of a bond of
type $b$ between a pair of parallel spins $u$ and $v$ will $u\oplus b\oplus v$,
rather than just $u\oplus v$ as in the original Ising model.  As part of our
kinematics---not as a peculiarity of an individual dynamics---we shall retain
the Ising prescription that a spin will flip if and only if such a move leaves
  %% PASTE HERE NEW TEXT
 the energy unchanged for a given assignment of the four surrounding bond
types. However, within this kinematics, we shall for the moment restrict our
attention to systems whose structural parameters (those determining a dynamics,
and in our case consisting of the assignment of bond types) be \emph{time- and
space-invariant}, that is, for each specific dynamics of our kinematics, the
behavior of a spin shall not depend on \emph{where} the spin is located and on
\emph{what time it is}, but only on the values of its neighbors.

With 
the
latter
restriction,
 we are left with as many \emph{global} assignments of structural parameters
as there are choices of \emph{local} assignments---that is, the 16 distinct
assignments of four bond types around a spin.  For each of these structures we
must tabulate, by running over all possible states, the corresponding dynamics
(a function from old global state to new global state) and the corresponding
energy function (a function from global state to an integer).  Observe that two
distinct structures may well end up yielding the same dynamics or the same
energy function.
 Our goal will be achieved if (a) no two dynamics are specified by the
\emph{same} energy function (in other words, the energy function
\emph{distinguishes} between dynamics), and (b) every dynamics has \emph{at
least one} energy function that specifies it. Since, by construction, energy is
additive down to the scale of individual four-bond ``patches,'' the local
``variational principle'' employed in \sect{generate} will then allow us not
only to indirectly denote, but to explicitly \emph{generate} the corresponding
vectorial-form (that is, state-to-state) dynamics.

This is indeed the case. Since we are presenting here an original approach, it
is important not to deceive ourselves on delicate aspects of an argument. For
this reason, as a ``sanity check'' we have explicitly tabulated dynamics and
energy function for all 16 elements of our generalized Ising kinematics on a
finite, toroidally wrapped-around model. (We used just a 4\by4 torus; though
small, this size is large enough to rule out spurious degeneracies. Full
scripts, output data, and statistics are available at
\url{www.ioc.ee/~silvio/nrg/}.) A formal proof is subsumed within the
properties of the numerical integration scheme of \fig{octa}, which constitute
a proof in the more general context of space- and time-dependent dynamics.  In
fact, in the next section we shall remove the constraint that the
structure---and the attendant dynamics---be spacetime uniform.

\Sect[spacetime]{Space- and time-dependent dynamics}

Energy in these generalized Ising models is a function of the global
state---a function ultimately specified by the structure parameters of
the specific model. If the parameters are allowed to change from one
step to the next, then this function is in turn a function \emph{of
time}. In fact, the type of the bond joining two given spins will
depend on ``what time it is'' at the moment of performing a transition
 $\langle q_\text{A},q_\text{B}\rangle\to\langle q_\text{B},q_\text{C}\rangle$.
 The ``new state'' $\langle q_\text{B},q_\text{C}\rangle$ at the end of this
transition reappears as the ``old state'' at the beginning of the following
transition,
 $\langle q_\text{B},q_\text{C}\rangle\to\langle q_\text{C},q_\text{D}\rangle$,
 but now it is a \emph{different time}, and the energy of the \emph{same state}
(of spins) has to be re-evaluated accordingly to the new assignment of
structural parameters.

This arrangement of things, which is essentially a numerical integration
scheme, is graphically illustrated by \fig{octa}. There, the integers along the
time axis represent discrete time instants, those along $x$ or $y$, discrete
spatial positions. A spin is represented by a thick vertical line, interrupted
by a conditional-permutation gate ($\oplus$) every time that it is the spin's
turn to be updated.
 \iffalse to each of them there is associated a configuration of spins,
alternately in even and odd spatial position (the absolute origin of this
parity alternation is immaterial).
 \fi
 A system state is an ordered pair of configurations associated with two
consecutive integer times. This state may conveniently be labeled with the
``half-time'' in between; that is, one may denote by $\frac12$ the state
consisting of the configurations at 0 and 1, and so forth. Steps of the
dynamics are nominally associated with integer values of time.  So, for
instance, state $1\frac12$ is the ``new state'' for step 1 but the ``old
state'' for step 2. Horizontal arcs of unit length (thin lines) represent
interspin bonds as at the time of a step; letters may be used to specify the
nature of a bond; so $r$ denotes the nature of the bond between spins 1 and 2
at step 1, while $s$ denotes the nature of the same bond at step 2.

 \kern-10pt
 \Fig[octa]{%
 \def\een{\begin{picture}(16,16)(-8,-8)
   \put(.1,-.1){\makebox(0,0){$\oplus$}}
   \put(-8,0){\line(1,0){4.5}}
   \put(8,0){\line(-1,0){4.5}}
   \thicklines
   \put(0,-8){\line(0,1){4.5}}
   \put(0,8){\line(0,-1){4.5}}
   \end{picture}}
 \def\odd{\begin{picture}(16,16)(-8,-8)
   \put(0,0){\circle*{3}}
   \put(-8,0){\line(1,0){16}}
   \put(8,0){\line(-1,0){4.5}}
   \thicklines
   \put(0,-8){\line(0,1){16}}
   \end{picture}}
 (a)\quad$\vcenter{\hbox{
 \begin{picture}(72,72)(-8,-8)
 %\put(0,0){\framebox(64,64){}}
  \put( 0,48){\odd}\put(16,48){\een}\put(32,48){\odd}\put(48,48){\een}
  \put( 0,32){\een}\put(16,32){\odd}\put(32,32){\een}\put(48,32){\odd}
  \put( 0,16){\odd}\put(16,16){\een}\put(32,16){\odd}\put(48,16){\een}
  \put( 0, 0){\een}\put(16, 0){\odd}\put(32, 0){\een}\put(48, 0){\odd}
  \footnotesize
  \put(32,43){\cstrip{s}}
  \put(32,27){\cstrip{r}}
  \put(8,-6){\cstrip{0}}\put(24,-6){\cstrip{1}}\put(40,-6){\cstrip{2}}\put(56,-6){\cstrip{3}}
  \put(-4,8){\cstrip{0}}\put(-4,24){\cstrip{1}}\put(-4,40){\cstrip{2}}\put(-4,56){\cstrip{3}}
  \put(-12,-6){$t\uparrow$}
  \put(64,-8){\shortstack{$x$\\$\rightarrow$}}
 \end{picture}}}$
 \quad\quad\quad\quad$\vcenter{\hbox{%
 \begin{picture}(56,72)(0,-4)
  \put(16,40){\makebox(0,0){$\oplus$}}
  \put(48,16){\makebox(0,0){$\oplus$}}
  \put(8,24){\makebox(0,0){$\scriptstyle\oplus$}}
  \put(40,64){\makebox(0,0){$\scriptstyle\oplus$}}
  \put(16,8){\circle*{3}}
  \put(48,48){\circle*{3}}
  \thicklines
  \put(16,0){\line(0,1){37}}\put(16,64){\line(0,-1){20.5}}
  \put(48,4){\line(0,1){9}}\put(48,60){\line(0,-1){40.5}}
  \thinlines
  \put(0,36){\line(4,1){13}}\put(56,50){\line(-4,-1){36.5}}
  \put(8,6){\line(4,1){38}}\put(61.3,19.35){\line(-4,-1){10}}
  \thicklines
  \put(8,16){\line(0,1){6}}\put(8,64){\line(0,-1){37.5}}
  \put(40,24){\line(0,1){37.5}}
  \thinlines
  \put(0,54){\line(4,1){38}}
  \put(-4,21){\line(4,1){10}}\put(56,36){\line(-4,-1){45.5}}
  \put(4,64){\line(1,-2){10.5}}\put(26,20){\line(-1,2){8.5}}
  \put(52,40){\line(-1,2){10.75}}
  \put(36,40){\line(1,-2){10.5}}\put(54.5,3){\line(-1,2){5}}
  \put(20,0){\line(-1,2){10.8}}\put(1.25,37.5){\line(1,-2){5.5}}
  \put(8,56){\circle*{2}}
  \put(40,32){\circle*{2}}
  \thicklines
  \put(24,16){\line(0,1){16}}
  \thinlines
  \put(19,22.75){\line(4,1){10}}
  \put(24,24){\circle*{3.5}}
  \footnotesize
  \put(-6,42){$t\uparrow$}
  \put(34,0){\shortstack{$x$\\$\rightarrow$}}
  \put(-1,2){\shortstack{\hbox{$y$}\\$\nwarrow$}}
 \end{picture}}}$\quad(b)%
}
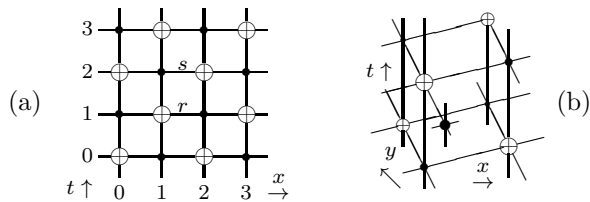
 {(a) Spacetime spin grid ($y$ not shown). (b) Detail of full grid, with both
$x$ and~$y$.}
 \kern-10pt

\noindent The value of a bond remains steady during spin updating (at integer
times) but may change between updatings (at half-integer times). On this
occasion, the energy of a state is accordingly re-evaluated. Thus, even if
energy is conserved during spin updating, in a time-dependent system the energy
may still change when the \emph{bonds} are altered.

Thus, it will generally happen that energy changes with time. With the present
\emph{full} class of dynamics (\ie\ one which allows arbitrary spacetime
variation of the structural parameters) of the given generalized Ising
kinematics, we arrive at the conclusion that energy is \emph{necessarily
conserved} \emph{if} the spacetime structure is time-invariant (because the
transitions themselves are by construction energy conserving, and with
time-invariant bond types no re-evaluation of the energy need occur). On the
other hand, when the spacetime structure is \emph{not} time invariant, \emph{at
least some} of the dynamics are not energy-conserving. We can thus rightfully
say that, in the present context, ``Ising energy'' \emph{as a whole} (not the
energy of a specific Ising dynamics) is the quantity that is conserved
\emph{because of} the time-invariance of the dynamics.

For this class of dynamics, then, in spite of the discreteness of the dynamics,
a non-frivolous transliteration of Noether's theorem applies with full force,
yielding energy conservation from time invariance.

\Sect[concl]{Conclusions}

We've shown that certain aspects of Noether's theorem apply to dynamical
systems beyond ordinary analytical mechanics, and \emph{why they do}. We were
able to shed the requirement that the dynamics be
\emph{continuous}; still, some form of \emph{second-order} discipline seems to
remain essential for a symmetry to give rise to a conservation law.
 \iffalse
 The quantity that has been called energy of an Ising spin
system not only is not changed by the dynamics unless the nature of
the bonds change with time, but also allows recovering the latter up
to an equivalence---which, in this case, is just the one which is
essential,
\ie, inducing the same one-step evolution.
 \fi
This is just the beginning of what promises to be a productive line of research.
 \iffalse As a next step we plan to turn our attention to conservation of
\emph{momentum} vs \emph{translational} invariance, and of angular momentum vs
rotational invariance.
 \fi
 \iffalse
 The more systems are considered, the more ``energies'' will be available as
experimental results. And other invariants are going to be sought, such as
\emph{momentum} which is associate to translational invariance. All these
quantities will contribute to the formulation of general principles, which are
expected to take the role of Noether's theorem for discrete systems; and on the
other hand, to the establishment of constraints to the validity of said
principles, such as the hypotheses from~\sect{noether} bound that of the
classical Noether's theorem.
 \fi

\end{document}

%% file: ttpaper.tex
 \input smallbib		% to cast the bibliography in \small type

\makeatletter
\long\def\proofbox#1{\gdef\@proofbox{#1}}
\proofbox{\small{\tt\@ifundefined{sourcepath}{}{\sourcepath/}\jobname}\\\today}

\def\proofref#1{\proofbox{\small{\tt#1\par
	[edited by Tom Toffoli for personal use]\par
	{\tt\sourcepath/\jobname}, 
	\number\month/\number\day/\number\year
	\par}}}

 \def\affil#1{\\{\small\sl#1\par}}
 \long\def\author#1{\gdef\@author{#1}}
 \author{Tommaso Toffoli ({\tt tt\char"40bu.edu})\affil{Electrical and
Computer Engineering, Boston University, Boston, MA 02215}}

 \long\def\abstract#1{\gdef\@abstract{#1}}
 \abstract{}

\def\leftfrac{.32}
\def\ritefrac{.68}
\def\@maketitle{\newpage\noindent\leavevmode
  \begin{minipage}[t]{\leftfrac\textwidth}
    \hrule height0pt
    \@proofbox
  \end{minipage}\hfil
  \begin{minipage}[t]{\ritefrac\textwidth}
    \hrule height0pt
    \raggedleft
    \LARGE\@title\par
    \vskip4pt
    \large\@author
  \end{minipage}
  \vskip8pt
  \ifx\@abstract\@empty\else{\vskip.5em\leftskip1.25in\parskip4pt\small\@abstract\par\vskip.5em}\fi
  \noindent
  \rule{\textwidth}{0.4pt}
  \vskip16pt}

\makeatother

%% file: latex+4.tex
\def\ifundefined#1{\expandafter\ifx\csname#1\endcsname\relax}

 \makeatletter
 \DeclareRobustCommand\em
        {\@nomath\em \ifdim \fontdimen\@ne\font >\z@
                       \upshape \else \slshape \fi}

\def\@begintheorem#1#2{\sl \trivlist \item[\hskip \labelsep{\bf #1\ #2}]}
\def\@opargbegintheorem#1#2#3{\sl \trivlist
     \item[\hskip \labelsep{\bf #1\ #2\ (#3)}]}

 \newenvironment{fineprint}{\par\medskip\begingroup\small}{\par\endgroup\medskip}

 \newcommand{\ie}{i.e.,}

 \newcommand{\cf}{cf.\ }

 \mathchardef\BY="0202
 \newcommand{\by}{\ensuremath{\BY}}              % 3\by4 will spell 3x4

 \def\@empty{}
 \newcommand{\asin}[2][]{{%extra braces for lower level, to forget \@cite
    \def\t@mp{#1}%
    \def\@cite##1##2{\marginpar{\hfil{\footnotesize$
    \ifx\t@mp\@empty\text{##2}\else\frac{\text{##1}}{\text{##2}}\fi$}\hfil}}%
\cite[#1]{#2}}}
 \let\asin\cite

 \def\pages#1{}

 \newcommand{\sectlabel}[1]{\label{sect:#1}}
 \newcommand{\footlabel}[1]{\label{foot:#1}}
 \newcommand{\eqlabel}[1]{\label{eq:#1}}
 \newcommand{\figlabel}[1]{\label{fig:#1}}
 
 \newcommand{\itmlabel}[1]{\label{itm:#1}}

 \newcommand{\Chapt}[2][]{\def\t@mp{#1}%
\chapter{#2} \ifx\t@mp\@empty\else\sectlabel{#1}\fi}
 \newcommand{\Sect}[2][]{\def\t@mp{#1}%
\section{#2} \ifx\t@mp\@empty\else\sectlabel{#1}\fi}
 \newcommand{\Subsect}[2][]{\def\t@mp{#1}%
\subsection{#2} \ifx\t@mp\@empty\else\sectlabel{#1}\fi}
 \newcommand{\Foot}[2][]{\def\t@mp{#1}%
\footnote{#2} \ifx\t@mp\@empty\else\footlabel{#1}\fi}
 \newcommand{\Itm}[2][]{\def\t@mp{#1}%
\item \ifx\t@mp\@empty\else\itmlabel{#1}\fi#2}
 \newcommand{\Eq}[2][]{\def\t@mp{#1}%
\begin{equation}#2\ifx\t@mp\@empty\notag\else\eqlabel{#1}\fi\end{equation}}
 \newcommand{\Eqaligned}[2][]{\def\t@mp{#1}%
\begin{equation}\begin{aligned}#2\end{aligned}
\ifx\t@mp\@empty\notag\else\eqlabel{#1}\fi
\end{equation}}
 \newcommand{\Eqmultline}[2][]{\def\t@mp{#1}%
\begin{multline}#2\ifx\t@mp\@empty\notag\else\eqlabel{#1}\fi
\end{multline}}
 \newcommand{\Eqgathered}[2][]{\def\t@mp{#1}%
\begin{equation}\begin{gathered}#2\end{gathered}
\ifx\t@mp\@empty\notag\else\eqlabel{#1}\fi
\end{equation}}

  \newcommand{\sect}[1]{\S\ref{sect:#1}}      % Ref. to section or subsection
 \newcommand{\eq}[1]{(\ref{eq:#1})}	% Ref. to equation
 \newcommand{\fig}[1]{Fig.\,\ref{fig:#1}}

 \def\thlabel#1{\label{th:#1}}

 \newcommand{\Theor}[2][]{\def\t@mp{#1}%
\begin{theorem}#2\ifx\t@mp\@empty\else\thlabel{#1}\fi\end{theorem}}
 \newcommand{\Lemma}[2][]{\def\t@mp{#1}%
\begin{lemm}#2\ifx\t@mp\@empty\else\thlabel{#1}\fi\end{lemm}}
 \newcommand{\Defi}[2][]{\def\t@mp{#1}%
\begin{definit}#2\ifx\t@mp\@empty\else\thlabel{#1}\fi\end{definit}}
 \newcommand{\Prop}[2][]{\def\t@mp{#1}%
\begin{proposition}#2\ifx\t@mp\@empty\else\thlabel{#1}\fi\end{proposition}}

 \long\def\endsubsection#1{\smallskip\hbox to\hsize{\leaders\hrule\hfill\ \sect{#1}}\medskip}

  \def\@arabic#1{\number #1} % my redefinition

\long\def\@makecaption#1#2{
	\vskip\abovecaptionskip
	\sbox\@tempboxa{{\small #1: #2}}%
	\ifdim\wd\@tempboxa>\hsize
	    {\small #1: #2\par}
	\else
	   \global\@minipagefalse
	   \hbox to\hsize{\hfil\box\@tempboxa\hfil}
	\fi
	\vskip\belowcaptionskip}

\def\figstrut#1{\hbox to\linewidth{\vrule height#1\hfill}}

\newcommand{\Fig}[3][]{% [label], picture, caption
\begin{figure}[!htb]
 \centering{\leavevmode#2}%
 \caption{#3}
 \figlabel{#1}
\end{figure}                 }

\def\cstrip#1{\setbox0=\hbox{$#1$}\kern-.5\wd0\lower2pt\box0}
\def\rstrip#1{\setbox0=\hbox{$#1$}\kern-\wd0\lower2pt\box0}
\def\lstrip#1{\setbox0=\hbox{$#1$}\lower2pt\box0}
\def\tstrip#1{\setbox0=\hbox{$#1$}\kern-.5\wd0\lower\ht0\box0}
\def\bstrip#1{\setbox0=\hbox{$#1$}\kern-.5\wd0\raise\ht0\box0}
\def\Lstrip#1{\setbox0=\hbox{$\mskip2mu#1$}\lower2pt\box0}

\def\icpad{\thinspace}
\def\codepad{.4em}
\def\codeskip{\kern1pt}
\let\codestop\relax

\def\ic{\icpad\begingroup \tt \let\do\@makeother \dospecials 
          \@ifstar{\@sic}{\@ic}}
\def\@sic#1{\def\@tempa ##1#1{##1\endgroup\icpad}\@tempa}
\def\@ic{\obeyspaces \frenchspacing \@sic}

\def\codeline{\par\codeskip\noindent\begingroup\leftskip=\codepad\tt \let\do\@makeother \dospecials 
          \@ifstar{\@scodeline}{\@codeline}}
\def\@scodeline#1{\def\@tempa ##1#1{##1\ \codestop\par\endgroup\codeskip}\@tempa}
\def\@codeline{\obeyspaces \frenchspacing \@scodeline}

\def\code{\@code\frenchspacing\@vobeyspaces\verbatim@start}
\def\@code{\the\every@verbatim
 \par\noindent
  \@beginparpenalty \predisplaypenalty
  \leftskip=\codepad\rightskip\z@
  \parindent\z@\parfillskip\@flushglue\parskip\z@
  \@@par
  \def\par{%
    \if@tempswa
      \leavevmode\null\@@par\penalty\interlinepenalty
    \else
      \@tempswatrue
      \ifhmode\@@par\penalty\interlinepenalty\fi
    \fi}%
  \def\@noitemerr{\@warning{No verbatim text}}%
  \obeylines
  \verbatim@font
  \let\do\@makeother \dospecials
  \everypar \expandafter{\the\everypar \unpenalty}}

\makeatother

\newcommand{\Codeline}[1][]{\def\t@mp{#1}%
\par\smallskip\noindent\begingroup\leftskip=\codepad\tt \let\do\@makeother \dospecials 
          \@ifstar{\@sCodeline}{\@Codeline}}
\def\@sCodeline#1{\def\@tempa ##1#1{##1\par\endgroup\smallskip}\@tempa}
\def\@Codeline{\obeyspaces \frenchspacing \@sCodeline}